# DOUBLE-WELL ULTRACOLD-FERMIONS COMPUTATIONAL MICROSCOPY: WAVE-FUNCTION ANATOMY OF ATTRACTIVE-PAIRING AND WIGNER-MOLECULE ENTANGLEMENT AND NATURAL ORBITALS

## Benedikt B. Brandt, Constantine Yannouleas and Uzi Landman*

School of Physics, Georgia Institute of Technology, Atlanta, Georgia 30332-0430, USA

(JULY 1, 2015)

## ABSTRACT

"Bottom-up" approaches to the many-body physics of fermions have demonstrated recently precise number and site-resolved preparations with tunability of interparticle interactions in single-well, SW, and double-well, DW, nano-scale confinements created by manipulating ultracold fermionic atoms with optical tweezers. These experiments emulate an analogue-simulator mapping onto the requisite microscopic hamiltonian, approaching realization of Feynmans' vision of quantum simulators that "will do exactly the same as nature". Here we report on exact benchmark configuration-interaction computational microscopy solutions of the hamiltonian, uncovering the spectral evolution, wave function anatomy, and entanglement properties of the interacting fermions in the entire parameter range, including crossover from a SW to a DW confinement and a controllable energy imbalance between the wells. We demonstrate attractive pairing and formation of repulsive, highly-correlated, ultracold Wigner molecules, well-described in the natural orbital representation. The agreement with the measurements affirms the henceforth gained deep insights into ultracold molecules and opens access to the size-dependent evolution of nano-clustered and condensed-matter phases and ultracold-atoms quantum information.

*Corresponding Author: uzi.landman@physics.gatech.edu

Ingress to the origins of complex physical phenomena often requires experiments whereby theories are tested or suggested through artificial manipulations of physical circumstances. During the past decade, a cornucopia of new tools have emerged resulting from the discovery and advancement of methods for the preparation and trapping of ultracold atomic gases, controlled tuning of the interparticle interactions (via magnetic manipulation of the Feshbach resonance), and the creation of synthetic gauge fields through atom-light interactions in optical lattices of varied geometries and topologies[1,2]. The remarkable pristine nature of these systems, and the exquisite level of control that can be exercised over them, brought forth a realization of Richard Feynman's vision[3] for the construction of physical quantum simulators, capable of an *exact* simulation, of systems or situations that are computationally or analytically intractable. Indeed, in the past several years we witnessed a surge of realizations of such exact simulations addressing diverse fields (see reviews in refs.1 and 2), including in particular the behavior of strongly interacting fermions where computations are precluded because of the "fermion sign problem."[4]. These systems range from high-Tc superconductivity[1,2], collosal magnetoresistance[5] and quantum Hall effects[2] to atomic frequency resonators[6], interferometry[7,8], matter wave gyroscopes[9] and the development of scalable quantum computers with neutral atoms[10,11].

Progress aiming at a "bottom-up" approaches to the many-body physics of fermions, has been demonstrated in recent efforts to deterministically prepare and measure few fermion systems in a SW[12] or DW[13] confinement created with the use of optical tweezers[12-14]. This experimental methodology differs in a substantial way from the "top-down" approach of most experiments with an optical lattice which is loaded by a large ensemble of trapped atoms cooled in an external trap. Here[12-14], the precise number and site-resolved preparation and the tunability of the nature (attraction or repulsion) and strength of their interactions open access to the molecular world and the size-dependent evolution of clustered and condensed-matter phases.

With the parameters chosen, the DW experiment[13] can be regarded as an emulation of an exact analogue simulator in the Feynman sense[3], mapping the two interacting fermion problem onto the *exact microscopic* hamiltonian [see Eq. (1) below]. Under the ultracold experimental conditions (T ~ $10^{-9}$ K) the de Broglie wave length, $\lambda_{dB} = h/(mv) = h/(2\pi mkT)^{1/2}$, of a $^6$Li atom is of the order of 24 μm and the interwell separation in the double well can be varied from zero to a couple of microns; consequently, the system of ultracold atoms in the DW confinement is found in the deep quantum regime.

Here we advance benchmark exact solutions for the microscopic hamiltonian describing the confined interacting fermions for the parameter ranges accessible in the experiment. These solutions uncover the evolution of the spectrum, wave-function anatomy, and entanglement properties of the two interacting fermions, demonstrating attractive pairing and formation, for strong interatomic repulsion, of highly correlated ultracold Wigner molecules (UCWM) which are well-described in the natural orbital representation; UCWM for bosonic atoms have been introduced in ref. [15]. We consider two DW (see below and the Supporting Information, SI) configurations: (1) a so-called "linear arrangement" (LA) where two quasi one-dimensional (1D) wells, connected by a barrier between them, are located on the same axis (x), and (2) a so-called "parallel arrangement" (PA) where the quasi 1D wells are oriented along two parallel lines in the y direction being separated by a barrier in the x-direction; tunneling between the wells occurs in the x-direction through the long sides of the wells (namely the sides that are along the y- axis). Case (1), the LA configuration, is described in detail in the main text of the article, and case (2), the PA configuration (see SI, Fig,. S1) is discussed and compared (along with the LA configuration) with the experimental results[13] in the SI (see Fig. S2). The remarkable agreement (See SI, Fig. S2) between the calculated results and the measurements[13], validates these novel theoretical and experimental methodologies and affirms the henceforth gained deep insights into fundamental aspects of the chemistry and physics of molecular and condensed-phase materials.

We begin with a brief statement of the many-body hamiltonian of the one-dimensional (1D) two-fermion DW system in the aforementioned LA configuration , expressed (with N =2 here) as a sum of the single-particle part $H(i)$ (defined in the SI)  and the two-particle contact interaction,

$$H = \sum_{i=1}^{N} H(i) + \sum_{i=1}^{N} \sum_{j>i}^{N} g\, \delta(x_i - x_j)\,, \tag{1}$$

where $x_i - x_j$ denotes the relative distance between the $i$ and $j$ fermions( e.g. $^6$Li atoms).
The external confining potential [in $H(i)$] that models the DW is based on a two-dimensional two-center-oscillator[16] (TCO, see SI) that allows for independent variation of both the interwell separation, $d$, and of the barrier height $V_b$. It further allows consideration of a tilt $\Delta$ between the left and right wells. The 1D character of our CI treatment is enforced through the requirement that only the zero-point motion in the $y$ direction is of any relevance (see SI).

Evolution of the many-body spectra, for wells of equal depth (i.e., $\Delta = 0$, see DW profiles in the inserts), is displayed in Fig. 1, showing the 8 lowest energy states for the entire interwell distance range, from the united atom [$d = 0$, Fig. 1(a)] to full dissociation of the two-particle Feschbach molecule [$d = 2$ $\mu$m, Fig. 1(c)]. Since the CI calculation preserves the total spin, the energy curves are labeled as singlets ($s$) or triplets ($t$); the parity of the many-body states (See SI) is also conserved ($\Delta = 0$), and thus the corresponding states are labeled also as $\pm$. Overall the evolution of the spectra reflects the splitting of the united atom into two wells. That is, a double degeneracy appears gradually and it fully develops for complete dissociation (Fig. 1(c)) where the eight curves in Fig. 1(a) regroup into four (five) curves in the repulsive (attractive) region, respectively.

The energy curves (in all panels of Fig. 1) fall into two groups: those that are independent of the interaction strength $g$ (horizontal lines) and those that depend on $g$. In all instances the energy of the triplet states is independent of $g$, as found also for a single well[17], due to the exchange hole imposed by the Pauli exclusion principle. The energies of the singlet states are dependent, in general, on $g$, except in the case of full dissociation when the singlet states having *one fermion in each well* become degenerate with corresponding triplet states (the exchange integral vanishes), see Fig. 1(c). For example, the lowest two degenerate horizontal lines in Fig. 1(c) correspond to Heitler-London-type (HL-type) wave functions (singlet and triplet) of the form $|L \uparrow R \downarrow> \pm |L \downarrow R \uparrow>$ in analogy with the stretched natural $H_2$ molecule (with $L$ and $R$ signifying the left and right wells and $\uparrow, \downarrow$ the two spin projections). Such states approximate the highly entangled Bell states[10,18]. The energy curves that show a $g$-dependence correspond to singlet states having *both fermions in the same well*. This is a consequence of the contact interaction which is not effective at the longer distances introduced by the inter-well separation.

For $\Delta = 0$, and for either repulsive ($-1/g < 0$) or attractive ($-1/g > 0$) interactions, the conservation of parity leads to the formation of highly entangled NOON states[19] of the form $|L \uparrow L \downarrow> \pm |R \uparrow R \downarrow>$. The pair of degenerate *first excited* states (blue and orange color) in the repulsive range ($-1/g < 0$) of Fig. 1(c) are such NOON states, representing repulsive (excited) bound states[20]. The pair of degenerate *ground* states (green and dark brown) in the attractive range ($-1/g > 0$) of Fig. 1(c) are also NOON states.

In the attractive range of Fig. 1(c) (complete well separation), anti-crossings appear between a couple of singlet-state curves. These anti-crossings are absent in the spectra of the united atom [Fig. 1(a)] and result from the non-separability of the center-of-mass and relative motions of the two fermions; these motions are separable for a single harmonic trap.

Results for a DW with an inter-well tilt $\Delta/h = 0.5$ kHz and separation $d = 2\mu m$ are displayed in Fig. 2; the parameters fall within the same range as those used in the experiments[13]. In addition to the energy spectra in the repulsive range $g > 0$ (for the attractive range, see below and the last figure) shown in center panel, we display the results of analysis of selected many-body wave functions (for different states and/or $g$ values), exhibiting their single-particle densities (SPDs, green surfaces) and spin-resolved conditional probability distributions[21] (CPDs, red surfaces); the letter labels (**a**, **b**, **c**,....) relate the surface plots to the corresponding points on the various energy curves (for the definitions of the SPD and CPD see Methods). The spin-resolved CPD gives the spatial probability distribution of finding a second fermion with spin projection $\sigma$ under the condition that another fermion is located (fixed) at $\boldsymbol{r}_0$ with spin projection $\sigma_0$; $\sigma$ and $\sigma_0$ can be either up (↑) or down (↓).

In the non-interacting limit (Fig.2, far-left of the $-1/g$ axis in Fig.2 ), the ground-state wave function consists of a single determinant formed by the up- and down-spin fermions occupying the lowest $1s$ space orbital in the left well, and as a result the SPD is localized on the left side of the plot in panel **a**. This state is denoted as $|L \uparrow L \downarrow>$; note that no NOON state is formed since the parity is not conserved for $\Delta \neq 0$, unlike the case for $\Delta = 0$ (Fig. 1). Following the increase in the ground-state energy with increasing $g$ (i.e., staying on the dark brown curve), an anti-crossing develops, associated with a resonance region in the vicinity of $U \sim \Delta$; this region is highlighted by a gray box in the energy plot of Fig. 2. This resonance (details displayed in Fig. 3) involves the singlet state $|L \uparrow L \downarrow>$ (with both fermions residing in the left well) and the singlet Heitler-London state $|L \uparrow R \downarrow> - |L \downarrow R \uparrow>$ (with one fermion in each well occupying the corresponding left/right $1s$ space orbitals).

Two main themes, pertaining to the structure of the many-body wave functions exhibited in Fig. 2 (**a-h**), emerge: (1) *both* fermions are localized either in the left or right well; see the cases **a**, **c**, **d**, **e**, and **g**, which involve both singlets, **a**, **c**, **d**, and **e**, and a triplet, g, and (2) each well contains one fermion. In the latter case, the wave functions can be approximated either with the singlet (panels b and f) or triplet (panel h) variants of the HL wave functions. The single-well space orbitals involved in the formation of the HL-type wave functions are not restricted only to the $1s$ left- and right-well orbitals, but may involve $1p$ orbitals of the individual wells (compare, e.g., **f** and **h**); the orbitals involved ($1s$, $1p$) are explicitly indicated as subscripts; 1s and 1p refers to the

1D states with zero and one node, respectively. We recall here that the HL wave functions involving one space orbital from each well faithfully approximate the highly entangled two-qubit Bell states.

Of particular interest are cases **d** and **e** with both fermions in the left well. Focusing first on the double-humped density in panel **e**, it is apparent that the underlying wave function cannot be approximated as $|L_s \uparrow L_s \downarrow>$ having an up-spin and a down-spin fermions occupying the same $1s$ space-orbital in the left well (as is the case in panel **a**). Rather, the double-humped density indicates that the two fermions (due to the large repulsion) localize and avoid each other, forming an UCWM. The displayed CPD in panel **e** further supports formation of a UCWM – indeed, placing the down-spin fermion at the position of the right hump (black down arrow) the distribution of the up-spin fermion (red surface) is found to be located away from the black arrow, with its maximum at the position of the second (left) density hump. The wave function of this UCWM (singlet) is well-approximated by the two-determinant HL form $|L_l \uparrow L_r \downarrow> - |L_l \downarrow L_r \uparrow>$, where the subscripts $l$ and $r$ indicate the left and right humps in the density (green surface) of panel **e**. The case in Panel **d** describes an incipient UCWM; the multi-determinantal nature of the wave function is a signature of a correlated state[21]. The predicted formation of Wigner molecules (WMs) made of cold atoms is a remarkable discovery. Indeed, WMs have been initially predicted theoretically[21 - 24] and subsequently found experimentally[21,25], for strongly interacting electrons in two-dimensional (2D) quantum dots (QDs) at semiconductor interfaces. More recently WMs have been found in other 2D QDs[26], clean carbon nanotubes[27], and for biexciton states in 3D QDs[28].

The discovery of Wigner molecules made of four fermions (electrons) in a double well confinement using full configuration interaction calculations, allowed us to establish the correspondence between strong Wigner molecules and Heisenberg spin chains[16]; for an earlier analysis of the spin structure of WMs in single harmonic well confinements (including a quasi-linear, 1D, case) see ref. 29. It was shown that the full WM wave function can be mapped into a pure spin function.[16, 29]

In the limit of $-1/g = 0$, the UCWM may reach the regime of fermionization of two distinguishable fermions, which has been most recently realized for two $^6$Li atoms confined within a *single* harmonic trap[30]. In this limit, the energy of the UCWM (singlet, blue curve) becomes degenerate with the energy of the triplet state (orange horizontal straight line). Note the similarity in the densities and CPDs between panel **g** (triplet with $S_z = 0$) and panel e (singlet UCWM).

A detailed analysis of the resonance region (highlighted by the square box in Fig. 2) is displayed in Fig. 3. As aforementioned (Fig. 2), this resonance corresponds to the anticrossing resulting from the interaction between two singlet states, and has the form

$$c_1|L \uparrow L \downarrow> \pm c_2(|L \uparrow R \downarrow> -|L \downarrow R \uparrow>)/2. \tag{2}$$

In Fig. 2 and Fig. 3, left (right) of the resonance, on the blue curve one has $c_1 < c_2$ $(c_1 > c_2)$, whereas on the dark brown curve $c_1 > c_2$ $(c_1 < c_2)$. At resonance $c_1 = c_2$. In agreement with the corresponding SPDs (green surfaces) in panels I and III, the probability ratio for finding a fermion in the left or right well at the resonance points (denoted as **a** and **c** in Fig. 3) is 3:1. Further corroboration that the structure of the many-body states at resonance is well approximated by Eq. (2) is provided by the spin-resolved CPDs in **a** and **c**. Indeed, in both cases, if one locates the down-spin fermion in the middle of the left well (see black arrow), the probability distribution (red surface) of the up-spin fermion extends in both wells, and the ratio of the volumes under its left/right parts is 2:1. In contrast, if one locates the down-spin fermion in the right well, the spin-up fermion is found only in the left well. It is pertinent to note that the horizontal energy curve (green) in Fig. 3 corresponds to the HL-type ($|L \uparrow R \downarrow> +|L \downarrow R \uparrow>)/2$, as is also corroborated through an inspection of the SPD and CPDs associated with the many-body wave function at the point specified by **b**.

***Quantifying entanglement using the von Neumann entropy as a measure, and the natural orbitals.*** The theory of entanglement in a two-qubit space is associated with the celebrated Bell states, used earlier in investigations of quantum information processes implemented with ultracold atoms in optical lattices[5]. The CI many-body wave functions, however, are associated with larger Hilbert spaces for which a quantitative measure of entanglement is the von Neumann entropy[21, 31] $S_{vN}$ defined as

$$S_{vN} = -\text{Tr}(\rho \log_2 \rho) + C \tag{3}$$

where $\rho$ is the single-particle density matrix (SPDM, see SI for details) and $C = -\log_2 N$ is a constant, yielding $S_{vN} = 0$ for an uncorrelated single-determinant state. In keeping with previous literature on two electrons in semiconductor quantum dots[21,30], base 2 logarithms are used.

For *two* fermions, the eigenvalues $p_j$ and the eigenvectors $\phi_j^{NO}(\boldsymbol{r})$ of the SPDM provide key information[31,32] concerning the anatomy of the many-body wave function. The wave functions $\phi_j^{NO}(\boldsymbol{r})$ are known as the natural orbitals (NOs), introduced by Löwdin[32]. For a singlet state it has been shown[32] that

$$\Phi^{CI}(\boldsymbol{r}, r') = \sum_{j=1}^{M} d_j\, \phi_j^{NO}(\boldsymbol{r})\phi_j^{NO}(r')(\alpha\beta' - \beta\alpha'), \tag{4}$$

with $d_j = \pm\sqrt{p_j}$; a similar expression applies for the triplet. In conjunction with $S_{vN}$, knowledge of the $p_j$'s and NOs determines fully the anatomy (and degree of entanglement) of the many-body wave function by specifying the minimal number $M$ of Slater determinants (referred to also as the Slater rank[18] of the many-body wave function) that gives the most rapid converged approximation to $\Phi^{CI}$ (see the analysis below regarding the bar plots in Fig. 4).

The entanglement entropy $S_{vN}$ for two $^6$Li atoms in a double well with $d = 2\ \mu m$ and $\Delta/h = 0.5$ kHz (the same parameters as in Fig. 2) is displayed in Fig. 4. Given that the allowed maximum value for the von Neumann entropy in our CI calculations is $\log_2(2K) - \log_2(2) = 6.13$ (we use a basis of $K = 70$ single-particle space orbitals), it is remarkable that the calculated values in Fig. 4 remain smaller than 1.3 in the repulsive range, and in particular in the regime of strong correlations, i.e., for $-\frac{1}{g} \to 0-$. This reflects formation of a Wigner molecule. $S_{vN} = 1$ for all the triplets, i.e., the von Neumann entropy curves for all triplet states in the double well collapse to the single horizontal line. We note that the dark brown and purple curves approach vanishing entropy as $-1/g \to -\infty$; this is natural because in the weak-repulsion regime ($g \to 0+$) they correspond to the single-determinant wave functions $|L\uparrow L\downarrow>$ (dark brown) and $|R\uparrow R\downarrow>$ (purple).

In contrast to the bounded values ($< 1.3$) of $S_{vN}$ for repulsive interaction, in the attractive region, all the $S_{vN}$ values associated with the singlet ground and excited states of a highly-correlated and tightly-bound dimer (see Fig. 4) tend to increase without bound in the limit of $-1/g \to 0+$. This indicates that the wave function of the tightly-bound attractive dimer consists effectively of a large number of Slater determinants [see the bar plot for the $p_j$'s in panel (a)]. Naturally, for weak interparticle attraction the CI wave function approaches a single Slater determinant having vanishing von Neumann entropy [see panel (b) in Fig. 4]. This behavior contrasts with the mostly-two-determinant states found by us for all cases in the repulsive regime. In particular, for the strongly-repulsive highly-correlated UCWM regime, the corresponding bar plot in (g) contrasts sharply with that in (a). Indeed in panel (g) two SPDM eigenvalues ($p_1 = 0.81$ and $p_2 = 0.18$) dominate; a third one is sufficiently small and can be neglected. Furthermore, for both repulsive and attractive interactions, we found that the HL-type (one fermion in each well) singlet states (bar plots not shown) approach the maximally entangled Bell states for increasing well separation; indeed the corresponding SPDM eigenvalues $p_1 = p_2 \to 1/2$, $p_j = 0$ for $j \geq 3$, as $d \to \infty$, while the entangled Bell states have the form ($| \uparrow \downarrow >$ $\mp | \downarrow \uparrow >)/\sqrt{2}$ with $S_{vN} = 1$. Due to this association with the Bell states, the HL-type states in the double well are a promising candidate for the implementation of quantum logic gates[10].

The above findings suggest that progress in achieving highly accurate solutions to systems described by many-body hamiltonians of interacting particles (involving contact, or other, e.g. Coulomb, interactions), particularly for circumstances of strong interparticle correlations, would involve the employment of basis functions made of natural orbitals (see, e.g., ref. 33). Note the resemblance between the shape of the profiles of the wave-functions along the $x$-axis, $\Psi_x(NO_1)$ and $\Psi_x(NO_2)$, and the CI-calculated density [Fig. 4(g)]. We also remark that construction of such NOs may be achieved without the need for prior CI calculations[31], for example through the iterative-NO method[34].

The Bell states play a crucial role in the theory of quantum information and quantum computation; they consist precisely of two determinants, having as a result a von Neumann entropy $S_{vN} = 1$; the Bell states are the maximally entangled states in the space of two qubits. On the other hand, the CI wave function gives an exact solution to the many-body problem, but it comprises in general many determinants. The above findings show that the von Neumann entropy provides a quantitative diagnostic tool for identifying the special many-body states that are close to the Bell states, i.e., when $S_{vN} \sim 1$. In this respect, it also provides a measure of the deviation from the ideal Bell state. When $S_{vN} \sim 1$, we stress again that there are only two dominant Slater determinants in the basis of natural orbitals.

The insights obtained here via computational microscopy probing of the wave-function anatomy and entanglement characteristics of two fermionic ultracold atoms in an isolated double-well confinement, in juxtaposition with the demonstrated benchmark experimental capability[2] to prepare and control such a system (with single-site addressability) provides the impetus for further explorations of more complex systems built from such building blocks. The theoretical methodology that we have introduced, which has been shown here to result in agreement with the experiments for the same range of DW confining parameters as chosen experimentally (see SI, in particular Fig. S2, for comparison of the results of calculations for two DW configurations with the experiments in reference 13), covers as well a broader parameter range than the one used in the experiments. Moreover, this methodology is also applicable to systems with a larger number of interacting atoms and complex confining geometries, including multiwells and arrangements in higher dimensions.

## Methods

**Many-body definitions of the SPD, CPD and SPDM.** The single-particle density (SPD) is the expectation value of the one-body operator

$$\rho(\boldsymbol{r}) = <\Phi_{N,q}^{\text{CI}}| \sum_{i=1}^{N} \delta\left(\boldsymbol{r} - \boldsymbol{r}_i\right)|\Phi_{N,q}^{\text{CI}}>, \tag{5}$$

where $|\Phi_{N,q}^{\text{CI}}>$ denotes the q-th many-body (N particles) CI wave function.

The spin-resolved two-point anisotropic correlation function is defined as

$$P_{\sigma\sigma_0}(\boldsymbol{r},\boldsymbol{r}_0) = <|\sum_{i\neq j}\delta\left(\boldsymbol{r} - \boldsymbol{r}_i\right)\delta(\boldsymbol{r}_0 - \boldsymbol{r}_j)\delta_{\sigma\sigma_i}\delta_{\sigma_0\sigma_j}|\Phi_{N,q}^{\text{CI}}>. \tag{6}$$

Using the normalization constant $N(\sigma,\sigma_0,\boldsymbol{r}_0) = \int P_{\sigma\sigma_0}(\boldsymbol{r},\boldsymbol{r}_0)d\boldsymbol{r}$, we further define a related spin-resolved conditional probability distribution (CPD) as

$$P_{\sigma\sigma_0}(\boldsymbol{r},\boldsymbol{r}_0) = P_{\sigma\sigma_0}(\boldsymbol{r},\boldsymbol{r}_0)/N(\sigma,\sigma_0,\boldsymbol{r}_0). \tag{7}$$

The single-particle density matrix (SPDM), $\rho$, is given by

$$\rho_{\nu\mu} = \frac{<\Phi^{CI}|a_{\mu}^{\dagger}a_{\nu}|\Phi^{CI}>}{\sum_{\mu}<\Phi^{CI}|a_{\mu}^{\dagger}a_{\mu}|\Phi^{CI}>}, \tag{8}$$

and it is normalized to unity, i.e., $\text{Tr}\rho = 1$. The Greek indices $\mu$ (or $\nu$) count the spin orbitals $\chi_{\mu}(\boldsymbol{r})$ that span the single-particle space (of dimension $2K$).

**Acknowledgements**

We acknowledge a communication by Professor S. Jochim concerning the setup of the experiment in Reference 13. We thank the Air Force Office for Scientific Research for the support of this work. Calculations were carried out at the GATECH Center for Computational Materials Science.

**Supporting Information Available:** The double well potential, the configuration interaction method, comparison with experiments for two different double well configurations. This material is available free of charge via the Internet at http://pubs.acs.org.

**Author contributions**

C.Y. & U.L. conceived the paper, B.B. B & C.Y. performed computations. B.B.B., C.Y. and U.L. analyzed the results. B.B.B, C.Y. and U.L. wrote the manuscript.

**Additional information**

Supporting information is available in the online version of the paper

**Competing financial interests:** The authors declare no competing financial interests.

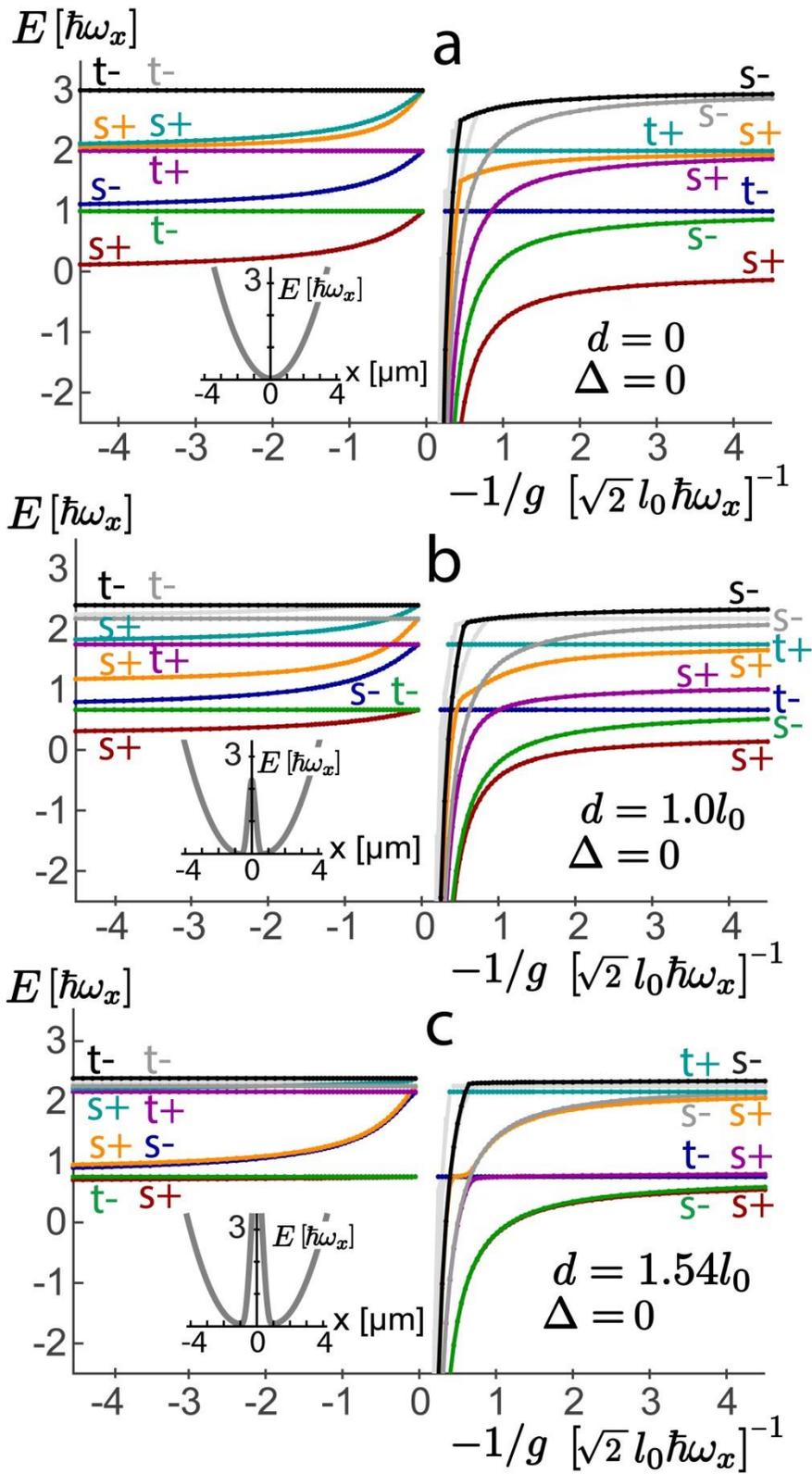

**Figure 1. Evolution of energy spectra of two fermions in a double well without tilt ($\Delta = 0$) as a function of the interaction strength $g$ and well separation $d$.** Both repulsive ($-1/g < 0$) and attractive ($-1/g > 0$) interparticle interactions are considered. The confining frequencies in the $x$ and $y$ directions are $\omega_x = 2\pi \times 1$ kHz and $\omega_y = 2\pi \times 100$ kHz, leading to an effective 1D confinement along the $x$ direction. In all three cases (a-c), the barrier heights $V_b$ (produced by the smooth neck) are given by $V_b = 18.18\, V_0$, where $V_0$ is the bare barrier of the TCO double well (see SI); $V_0/h = 0.125$ kHz and 0.297 kHz for a, b and c, respectively. This factor leads to strong anharmonicities in the confining double-trap potentials. The interwell separation is (a) $d = 0$, the "united atom" (single well), (b) $d = 1.297$ $\mu$m $= l_0$, and (c) $d = 2$ $\mu$m $= 1.543 l_0$, representing two rather well-separated wells, with ($l_0 \equiv l_{0x} = \sqrt{\hbar/(M\omega_x)}$) being the (left or right) harmonic-oscillator length. The mass corresponds to ultracold $^6$Li atoms, $M = 9.99\ 10^{-27}$ kg. The DW parameters in (c) are within the range of those used in the experiments[13]. The colors of the energy curves are consistent in all three panels. The horizontal curves in (c) correspond to HL-type (one fermion in each well) states that relate to the maximally spin-entangled two-qubit Bell states. Due to parity conservation, the $g$-dependent, doubly-degenerate first-excited (dark blue and orange) energy curves in the repulsive regime in (c) correspond to highly space-entangled NOON states of the form $(|2,0> \pm |0,2>)/\sqrt{2}$.

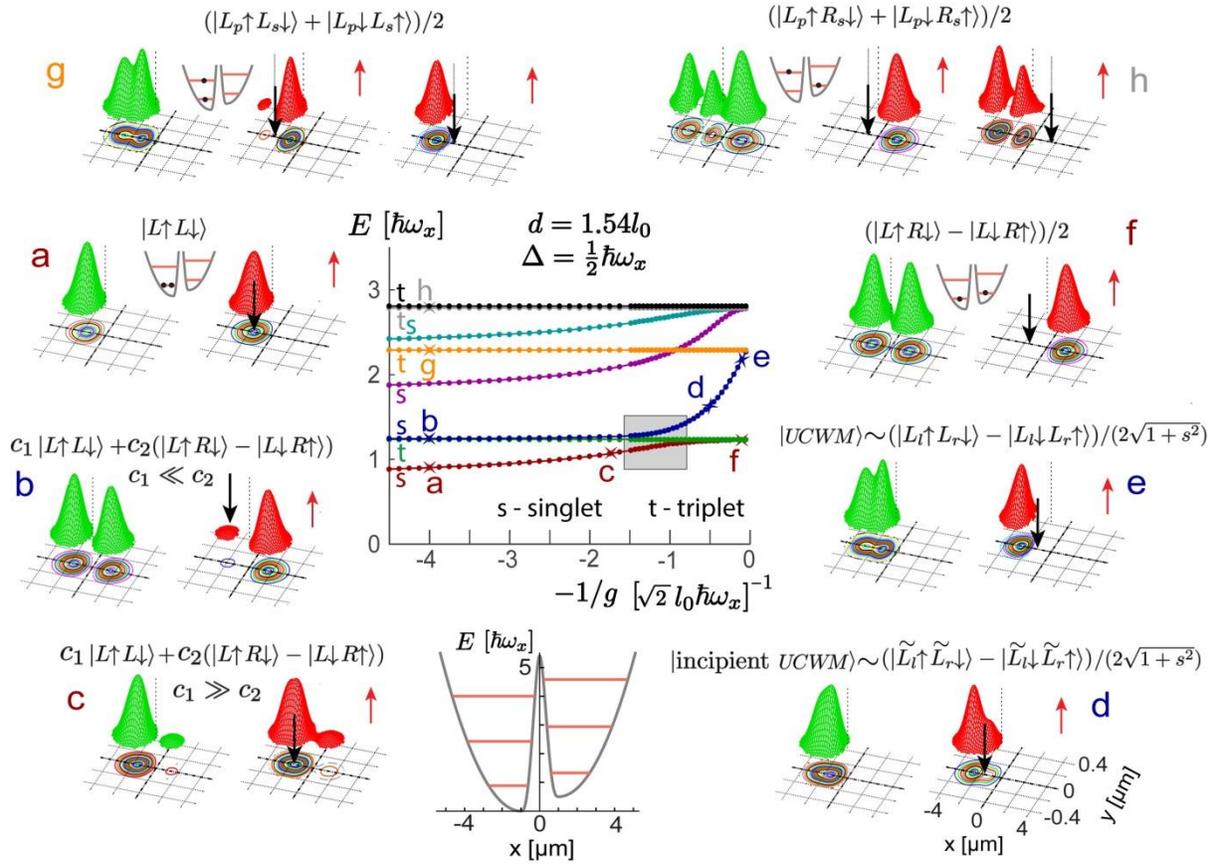

**Figure 2**

**Figure 2. Double well with a finite tilt $\Delta/h = 0.5$ kHz and well separation $d = 2 \mu m$ = $1.543l_0$.** The spectrum as a function of the strength of the interaction $g$ is displayed for the repulsive regime ($-1/g < 0$); for the spectrum for the attractive regime, see Fig. 4. The many-body wave-function anatomy (single-particle densities, SPDs, green surfaces, and spin-resolved conditional probability distributions, CPDs, red surfaces, see Methods) are illustrated for several representative instances, marked by letters a – h on the energy curves. The spin-resolved CPD gives the spatial probability distribution of finding a second fermion with spin projection $\sigma$ under the condition that another fermion is located (fixed) at $r_0$ with spin projection $\sigma_0$; $\sigma$ and $\sigma_0$ can be either up (↑) or down (↓), and in Fig. 2 the black arrow denoted the location of the observation point $r_0$ with $\sigma_0 = \downarrow$, and the red surface gives the probability distribution of the electron with spin ↑ (indicated by the red up-arrow). The abscissa values associated with the points a-h: $-1/g = -4/(\sqrt{2}l_0\hbar\omega_x)$ at a, b, g and h; $-1/g = -1.70/(\sqrt{2}l_0\hbar\omega_x)$ at c; $-1/g = -0.5/(\sqrt{2}l_0\hbar\omega_x)$ at d; $-1/g = -0.1/(\sqrt{2}l_0\hbar\omega_x)$ at e and f. The far-left part of the $-1/g$ axis represents the non-interacting limit. For $-1/g \rightarrow -\infty$, the ground state (brown curve) consists of two fermions in the left well (see panel a); the space-entangled NOON states (see caption of Fig. 1) do not survive a finite tilt. However, even in the non-interacting limit, the HL-type Bell-like states (with one fermion in each well) do survive the influence of the tilt with small modifications; see the density and CPD for the singlet state in panel b (point b is located on the blue energy curve). Increasing $g$ brings one to a resonance between the doubly-occupied singlet state in the left well and the HL-type singlet state discussed above. This resonance corresponds to an anticrossing region centered at $-1/g = -1.22/(\sqrt{2}l_0\hbar\omega_x)$ and is highlighted by a square. For strong repulsion, the two fermions minimize their interaction energy by avoiding each other, leading to the formation of a UCWM; note in panel e the two-humped density and the behavior of the CPD. The cases of two triplet states with spin projection $S_z = 0$ are elaborated in panels g and h; they have the structure of Bell states (↑↓> + ↓↑>)/$\sqrt{2}$. We have checked that the purple curve (associated SPDs and CPDs not shown) corresponds to both fermions being trapped in the right well. The labels $L$ and $R$ correspond to space orbitals localized on the left and right wells. The subscripts $l$ and $r$ denote space orbitals partially localized on the left and right side of a given well. The subscripts $s$ and $p$ denote $1s$-type and $1p$-type orbitals in the left or right well. The symbol $s$ in $2\sqrt{1 + s^2}$ denotes the overlap of left and right space orbitals comprising the singlet states. The confining frequencies in the $x$ and $y$ directions are $\omega_x = 2\pi \times 1$ kHz and $\omega_y = 2\pi \times 100$ kHz and $V_b = 18.18V_0$;

$V_b$ is measured from the bottom of the left well and $V_0/h = 0.297$ kHz. The mass corresponds to ultracold ${}^6$Li atoms, $M = 9.99 \ 10^{-27}$ kg. All chosen parameters are within the range of a recently reported experiment[13].

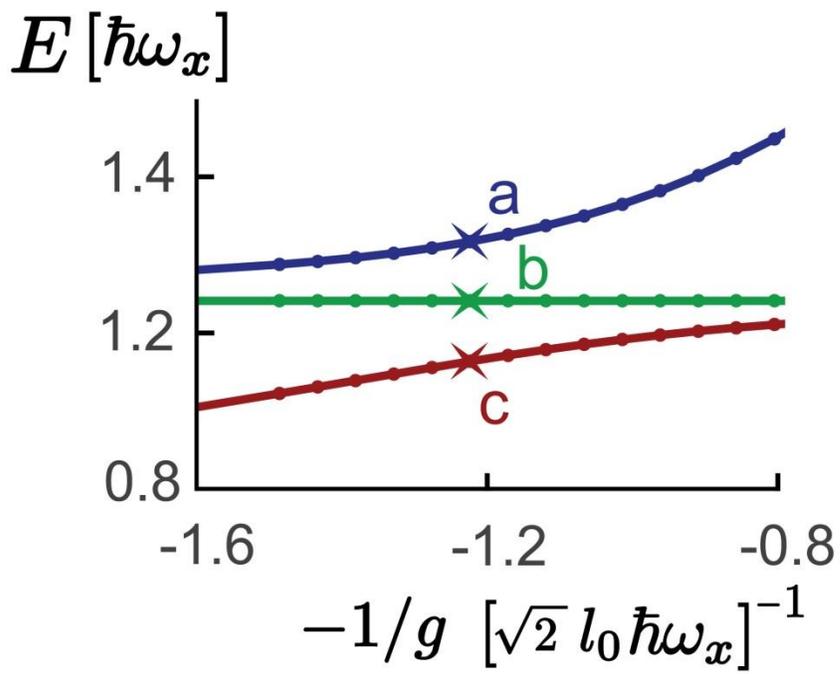

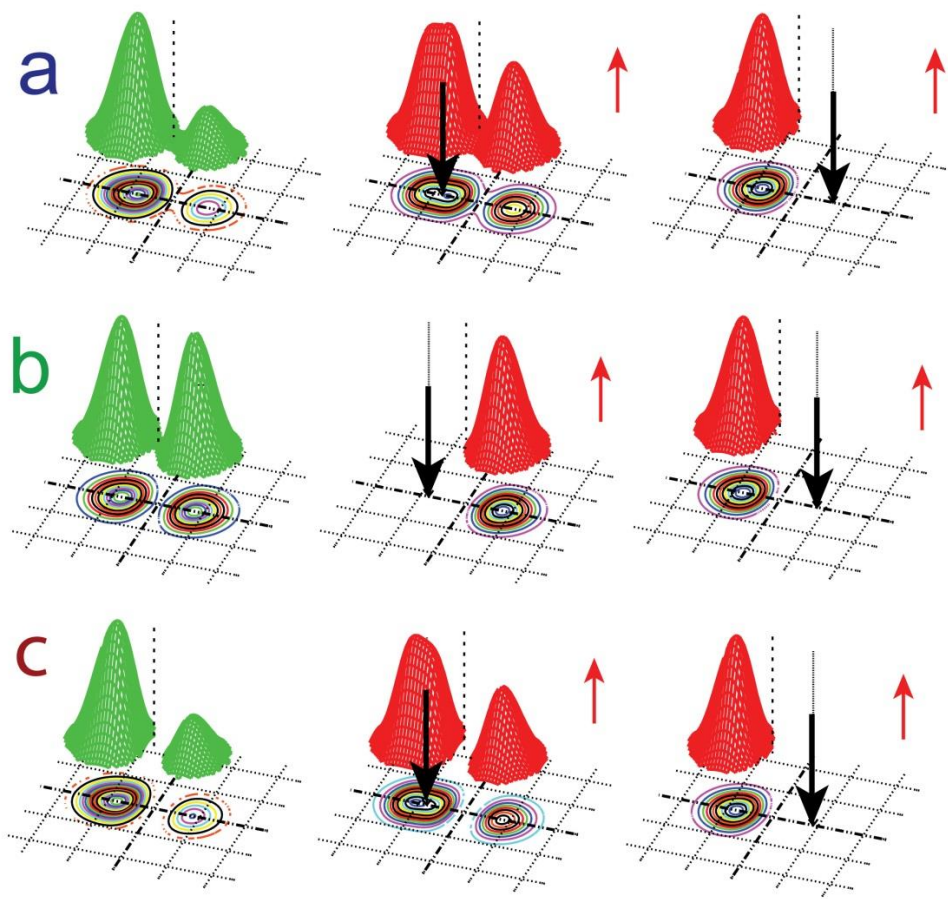

**Figure 3**

**Figure 3. The resonance region in a double well with a finite tilt $\Delta/h = 0.5$ kHz and well separation $d = 2\ \mu\text{m} = 1.543 l_0$.** Top: magnification of the anticrossing spectrum region enclosed in the square highlighted in Fig. 2. The resonance region is centered at $-1/g = -1.22/(\sqrt{2}l_0 \hbar \omega_x)$. At this value of $-1/g$, two labels correspond to the two singlets states (a, blue and c, brown) participating in the resonance $|RES^{\pm}> = c_1|L \uparrow L \downarrow> \pm c_2(|L \uparrow R \downarrow> - |L \downarrow R \uparrow>)/2$, namely the resonance between a state with both fermions in the left well and a state with one fermion in each well; it occurs when the repulsive interaction energy $U$ of the two fermions in the left well equals the tilt energy $\Delta$. A third label b corresponds to a triplet state (green) of the HL-type, $|TRI> = (|L \uparrow R \downarrow> + |L \downarrow R \uparrow>)/2$. Panels a, b, and c display the corresponding SPDs (green surfaces) and CPDs (red surfaces), supporting the intuitive expressions for the many-body wave functions, $|RES^{\pm}>$ and $|TRI>$, given above. Exactly at resonance $c_1 = c_2$; see text for a detailed description. The confining frequencies in the $x$ and $y$ directions are $\omega_x = 2\pi \times 1$ kHz and $\omega_y = 2\pi \times 100$ kHz and $V_b = 18.18 V_0$; $V_b$ is measured from the bottom of the left well and $V_0/h = 0.297$ kHz. The mass corresponds to ultracold $^6$Li atoms, $M = 9.99\ 10^{-27}$ kg. All chosen parameters are within the range of a recently reported experiment[13].

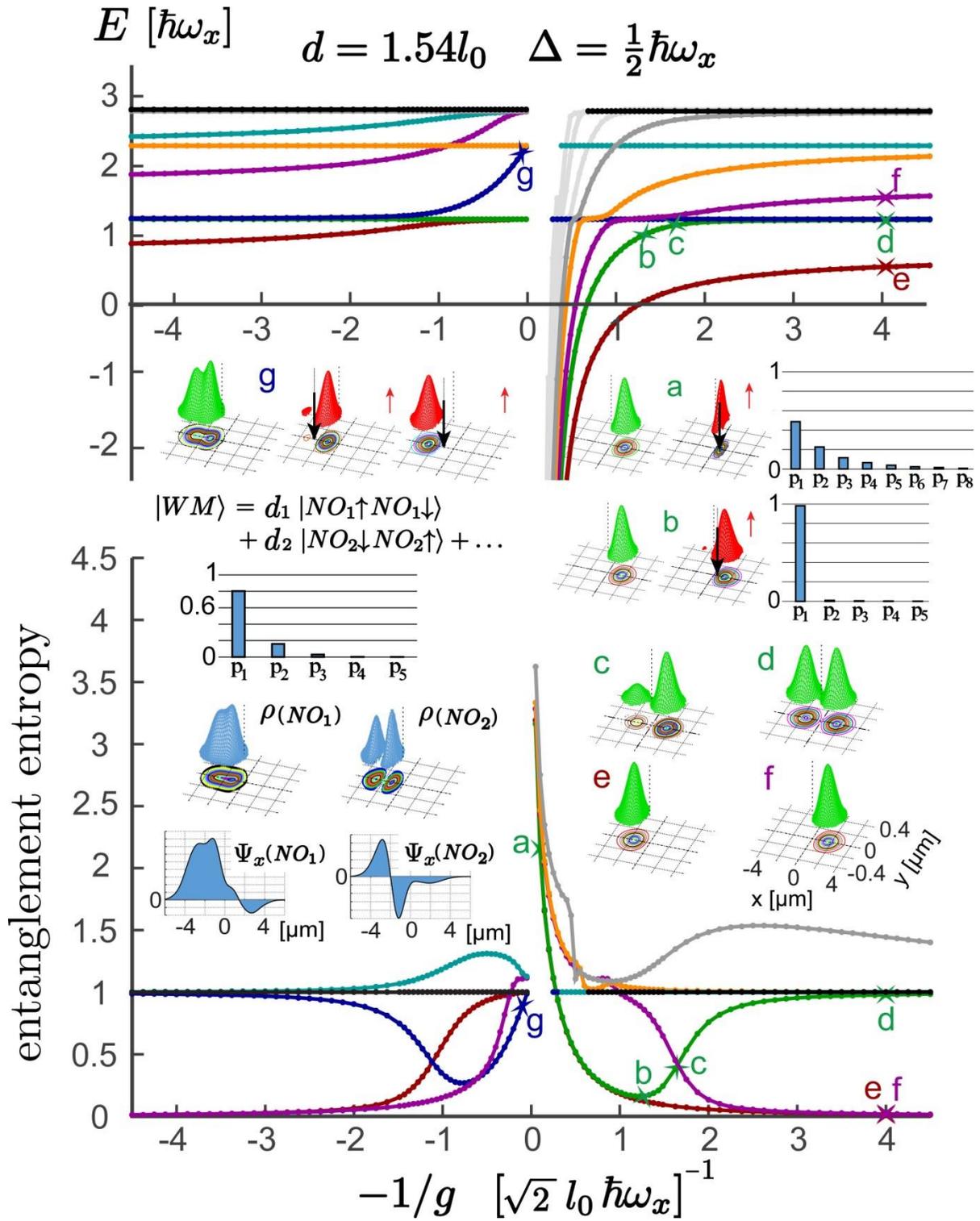

$E$ $[\hbar\omega_x]$    $d = 1.54 l_0$    $\Delta = \frac{1}{2}\hbar\omega_x$

$|WM\rangle = d_1\,|NO_1{\uparrow}NO_1{\downarrow}\rangle$
$\qquad + d_2\,|NO_2{\downarrow}NO_2{\uparrow}\rangle + \cdots$

$\rho(NO_1)$    $\rho(NO_2)$

$\Psi_x(NO_1)$    $\Psi_x(NO_2)$

entanglement entropy

$-1/g$    $\left[\sqrt{2}\,l_0\,\hbar\omega_x\right]^{-1}$

**Figure 4**

**Figure 4. Evolution of entanglement in a double well with a finite tilt $\Delta/h = 0.5$ kHz and well separation $d = 2\ \mu$m $= 1.543 l_0$.** The von Neumann entropy (bottom), in conjunction with the energy spectrum (top), is shown as a function of the strength of the inter-particle interaction strength $g$. Both repulsive $(-1/g < 0)$ and attractive $(-1/g > 0)$ interparticle interactions are considered. In addition to the single-particle densities (SPDs, green surfaces) panels (a)−(g), associated with the points marked by these letters on the energy and entropy curves, display (one or both of) the following quantities: conditional probability distributions (CPDs, red surfaces) and the single particle density matrix eigenvalues (blue bars) associated with the points a−g, marked on the energy and entropy curves. The point g on the repulsive side corresponds to the singlet-spin UCWM discussed previously in Fig. 2. For this case, the two-dominant (see the corresponding bar plot) natural orbitals are also plotted. As seen from both the orbital density, $\rho$, and the orbital wave-function cut, $\Psi_x$, along the $x$-axis, the zero-node $NO_1$ is two-peaked in contrast to the single-peak $1s$ orbital familiar from the non-interacting case. At the points labeled e (with both fermions residing in the left well, see associated densities) and f (with both fermions residing in the right well), the entropy is vanishing indicating a single-determinant wave function appropriate for the noninteracting limit. At the point d, the entropy is unity, indicating a maximally entangled two-determinant wave function of the HL-type (with one fermion in each well). At point b, the entropy is $\sim 0.2$, and the many-body wave function consists mainly of a single determinant [see the bar plot in panel (b)]. However, for strong attraction (point a), the number of Slater determinants grows out of bound [see the bar plot in panel (a)], a behavior that contrasts with that of the strong-repulsion UCWM case [compare bar plots in (a) and (g)]. For both left and right wells, $\omega_x = 2\pi \times 1$ kHz and $\omega_y = 2\pi \times 100$ kHz. The interwell barrier height $V_b = 18.18 V_0$;

$V_b$ is measured from the bottom of the left well and $V_0/h = 0.297$ kHz. The mass corresponds to ultracold $^6$Li atoms, $M = 9.99\ 10^{-27}$ kg. All chosen parameters are within the range of a recently reported experiment[13].

# SUPPORTING INFORMATION

## DOUBLE-WELL ULTRACOLD FERMIONS COMPUTATIONAL-MICROSCOPY: WAVE-FUNCTION ANATOMY OF ATTRACTIVE-PAIRING AND WIGNER-MOLECULES ENTANGLEMENT AND NATURAL ORBITALS

Benedikt B. Brandt, Constantine Yannouleas and Uzi Landman*
School of Physics, Georgia Institute of Technology, Atlanta, Georgia 30332-0430, USA

*Corresponding Author: uzi.landman@physics.gatech.edu

**1. Two-center-oscillator confining potential.** Following the recent experimental advances[1-3], and in particular those in ref. 2, we investigate here the quantum mechanical properties of two interacting fermionic ultracold atoms confined in a double well. (DW). . We consider two DW configurations: (1) a so-called "linear arrangement" (LA) where two quasi one-dimensional (1D) wells (see below Eq. S1), connected by a barrier between them, are located on the same axis (x), and (2) a so-called " parallel arrangement" (PA) where the quasi 1D wells are oriented along two parallel lines in the y direction being separated by a barrier in the x-direction; tunneling between the wells occurs in the x-direction through the long sides of the wells (namely the sides that are along the y- axis). Case (1), the LA configuration, is described in detail in the main text of the article, and case (2), the PA configuration, is discussed (see caption to Fig. S1 below) and compared (along with the LA configuration) with the experimental results[2], see Fig. S2 below.

To model the two interacting fermionic ultracold atoms confined in a double well we use a 2D many-body problem (as described below). In the LA configuration we enforce the 1D character by requiring that the trap confinement in the $y$-direction is much stronger than that in the $x$-direction (i.e. $\omega_y / \omega_{x \gg 1}$), with the result that only the zero-point motion in the $y$-direction is included in the calculations, whereas in the PA configuration we choose $\omega_y / \omega_x < 1$ (see the caption of Fig. S1 below).

In the 2D two-center-oscillator (TCO), the single-particle levels associated with the confining potential are determined by the single-particle hamiltonian[4,5]

$$H = \frac{\boldsymbol{p}^2}{2m} + \frac{1}{2}m\omega_y^2 y^2 + \frac{1}{2}m\omega_{xk}^2 x_k'^2 + V_{neck}(x) + h_k,　\quad\quad (S1)$$

where $x_k' = x - x_k$ with $k = 1$ for $x < 0$ (left well) and $k = 2$ for $x > 0$ (right well), and the $h_k$'s control the relative well-depth, with the tilt being $\Delta = h_2 - h_1$. $y$ denotes the coordinate perpendicular to the inter-dot axis ($x$). The most general shapes described by $H$ are two semiellipses connected by a smooth neck [$V_{neck}(x)$]; $x_1 < 0$ and $x_2 > 0$ are the centers of these semiellipses, $d = x_2 - x_1$ is the interdot distance, and $m$ is the atom mass.

For the smooth neck between the two wells, we use $V_{neck}(x) = \frac{1}{2}m\omega_{xk}^2[C_k x_k'^3 + D_k x_k'^4]\theta(|x| - |x_k|)$, where $\theta(u) = 0$ for $u > 0$ and $\theta(u) = 1$ for $u < 0$. The four constants $C_k$ and $D_k$ can be expressed via two parameters, as follows: $C_k = (2 - 4\epsilon_k^b)/x_k$ and $D_k = (1 - 3\epsilon_k^b)/x_k^2$, where the barrier-control parameters $\epsilon_k^b = (V_b - h_k)/V_{0k}$ are related to the actual height of the bare interdot barrier ($V_b$) between the two wells, and $V_{0k} = m\omega_{xk}^2 x_k^2/2$ (for $h_1 = h_2$, $V_{01} = V_{02} = V_0$).

The single-particle levels of $H$ are obtained by numerical diagonalization in a (variable-with-separation) basis consisting of the eigenstates of the auxiliary hamiltonian:

$$H_0 = \frac{\boldsymbol{p}^2}{2m} + \frac{1}{2}m\omega_y^2 y^2 + \frac{1}{2}m\omega_{xk}^2 x_k'^2 + h_k \ .　\quad\quad (S2)$$

The eigenvalue problem associated with the auxiliary hamiltonian (Eq. S2) is separable in $x$ and $y$, i.e., the wave functions are written as

$$\varphi_i(x,y) = X_\mu(x)Y_n(y),　\quad\quad (S3)$$

with $i \equiv \{\mu, n\}$, $i = 1, 2, \ldots, K$. The $Y_n(y)$ are the eigenfunctions of a 1D oscillator, and the $X_\mu(x \le 0)$ or $X_\mu(x > 0)$ can be expressed through the parabolic cylinder functions $U[\gamma_k, (-1)^k \xi_k]$, where $\xi_k = x_k'\sqrt{2m^*\omega_{xk}/\hbar}$, $\gamma_k = (-E_x + h_k)/(\hbar\omega_{xk})$, and $E_x = (\mu + 0.5)\hbar\omega_{x1} + h_1$ denotes the $x$-eigenvalues. The matching conditions at $x = 0$ for the left and right domains yield the $x$-eigenvalues and the eigenfunctions $X_\mu(x)$. The $n$ indices are integer. The number of $\mu$ indices is finite; however, they are in general real numbers.

**2. The configuration-interaction method.** As aforementioned, we use the method of configuration Interaction for determining the solution of the many-body problem specified by the

Hamiltonian (Eq. S1).

In the CI method, one writes the many-body wave function $\Phi_N^{\text{CI}}(\boldsymbol{r}_1, \boldsymbol{r}_2, \ldots, \boldsymbol{r}_N)$ as a linear superposition of Slater determinants $\Psi^N(\boldsymbol{r}_1, \boldsymbol{r}_2, \ldots, \boldsymbol{r}_N)$ that span the many-body Hilbert space and are constructed out of the single-particle *spin-orbitals*

$$\chi_j(x, y) = \varphi_j(x, y)\alpha, \quad \text{if} \quad 1 \leq j \leq K, \tag{S4}$$

and

$$\chi_j(x, y) = \varphi_{j-K}(x, y)\beta, \quad \text{if} \quad K < j \leq 2K, \tag{S5}$$

where $\alpha(\beta)$ denote up (down) spins. Namely

$$\Phi_{N,q}^{\text{CI}}(\boldsymbol{r}_1, \ldots, \boldsymbol{r}_N) = \sum_I C_I^q \, \Psi_I^N(\boldsymbol{r}_1, \ldots, \boldsymbol{r}_N), \tag{S6}$$

where

$$\Psi_I^N = \frac{1}{\sqrt{N!}} \begin{vmatrix} \chi_{j_1}(\boldsymbol{r}_1) & \cdots & \chi_{j_N}(\boldsymbol{r}_1) \\ \vdots & \ddots & \vdots \\ \chi_{j_1}(\boldsymbol{r}_N) & \cdots & \chi_{j_N}(\boldsymbol{r}_N) \end{vmatrix}, \tag{S7}$$

and the master index $I$ counts the number of arrangements $\{j_1, j_2, \ldots, j_N\}$ under the restriction that $1 \leq j_1 < j_2 < \cdots < j_N \leq 2K$. Of course, $q = 1, 2, \ldots$ counts the excitation spectrum, with $q = 1$ corresponding to the ground state. In our CI calculations full convergence is reached through the use of a basis of up to 70 TCO single-particle states; the TCO single-particle states automatically adjust to the separation $d$ as it varies from the limit of the unified atom $d = 0$ to that of the dissociation of the dimer (for sufficiently large $d$).

The many-body Schrödinger equation $H\Phi_{N,q}^{\text{EXD}} = E_{N,q}^{\text{EXD}} \Phi_{N,q}^{\text{EXD}}$ transforms into a matrix diagonalizatiom problem, which yields the coefficients $C_I^q$ and the eigenenergies $E_{N,q}^{\text{CI}}$. Because the resulting matrix is sparse, we implement its numerical diagonalization employing the well known ARPACK solver[6].

The matrix elements $< \Psi_N^I | H | \Psi_N^J >$ between the basis determinants [see Eq. (S7)] are calculated using the Slater rules[7]. Naturally, an important ingredient in this respect are the two-body matrix elements of the contact interaction,

$$g_{\text{2D}} \int_{-\infty}^{\infty} \int_{-\infty}^{\infty} d\,\boldsymbol{r}_1 d\boldsymbol{r}_2 \varphi_i^*(\boldsymbol{r}_1)\varphi_j^*(\boldsymbol{r}_2)\delta(\boldsymbol{r}_1 - \boldsymbol{r}_2)\varphi_k(\boldsymbol{r}_1)\varphi_l(\boldsymbol{r}_2), \tag{S8}$$

in the basis formed out of the single-particle spatial orbitals $\varphi_i(\boldsymbol{r})$, $i = 1, 2, \ldots, K$ [Eq. (S7)]. In our approach, these matrix elements are determined numerically and stored separately. The corresponding 1D interparticle interaction strengths, $g$, are extracted from $g_{2D}$ as follows

$$g = g_{2D} \int_{-\infty}^{\infty} d\,u[W(u)]^4, \tag{S9}$$

where $u$ is a dummy variable and $W$ is the lowest-in-energy single-particle state in the $y$ ($x$) direction for the LA (PA) configurations, respectively. In the LA configuration, $W$ coincides with $Y_0$, whereas in the PA configuration $W$ is a linear superposition of $X_\mu$'s due to the effect of the smooth neck.

The Slater determinants $\Psi_I^N$ [see Eq. (S7)] conserve the third projection $S_z$, but not the square $\hat{\boldsymbol{S}}^2$ of the total spin. However, because $\hat{\boldsymbol{S}}^2$ commutes with the many-body Hamiltonian, the CI solutions are automatically eigenstates of $\hat{\boldsymbol{S}}^2$ with eigenvalues $S(S+1)$. After the diagonalization, these eigenvalues are determined by applying $\hat{\boldsymbol{S}}^2$ onto $\Phi_{N,q}^{CI}$ and using the relation

$$\hat{\boldsymbol{S}}^2 \Psi_I^N = \left[ (N_\alpha - N_\beta)^2/4 + N/2 + \sum_{i<j} \varpi_{ij} \right] \Psi_I^N, \tag{S9}$$

where the operator $\varpi_{ij}$ interchanges the spins of fermions $i$ and $j$ provided that their spins are different; $N_\alpha$ and $N_\beta$ denote the number of spin-up and spin-down fermions, respectively. When $h_1 = h_2$ ($\Delta = 0$), the $xy$-parity operator associated with reflections about the origin of the axes is defined as

$$\hat{P}_{xy} \Phi_{N,q}^{CI}(\boldsymbol{r_1}, \boldsymbol{r_2}, \dots, \boldsymbol{r_N}) = \Phi_{N,q}^{CI}(-\boldsymbol{r_1}, -\boldsymbol{r_2}, \dots, -\boldsymbol{r_N}) \tag{S10}$$

and has eigenvalues $\pm 1$. With the two-center oscillator *Cartesian* basis that we use [see Eq. S7)], it is easy to calculate the parity eigenvalues for the Slater determinants, Eq. (S7), that span the many-body Hilbert space. Because $X_\mu(x)$ and $Y_n(y)$ conserve the partial $\hat{P}_x$ and $\hat{P}_y$ parities, respectively, one finds:

$$\hat{P}_{xy} \Psi_I^N = (-)^{\sum_{i=1}^{N}(m_i + n_i)} \Psi_I^N, \tag{S11}$$

where $m_i$ and $n_i$ count the number of single-particle states associated with the bare two-center oscillator [see the auxiliary Hamiltonian $H_0$ in Eq. (S2)] along the $x$ axis and the simple oscillator along the $y$ direction (with the assumption that the lowest states have $m = 0$ and $n = 0$, since they are even states). We note again that the index $\mu$ in Eq. S3 is not an integer in general, while $m$ here is indeed an integer (since it counts the number of single-particle states along the $x$ direction).

**3. Results for the double well parallel arrangement (PA) configuration**.

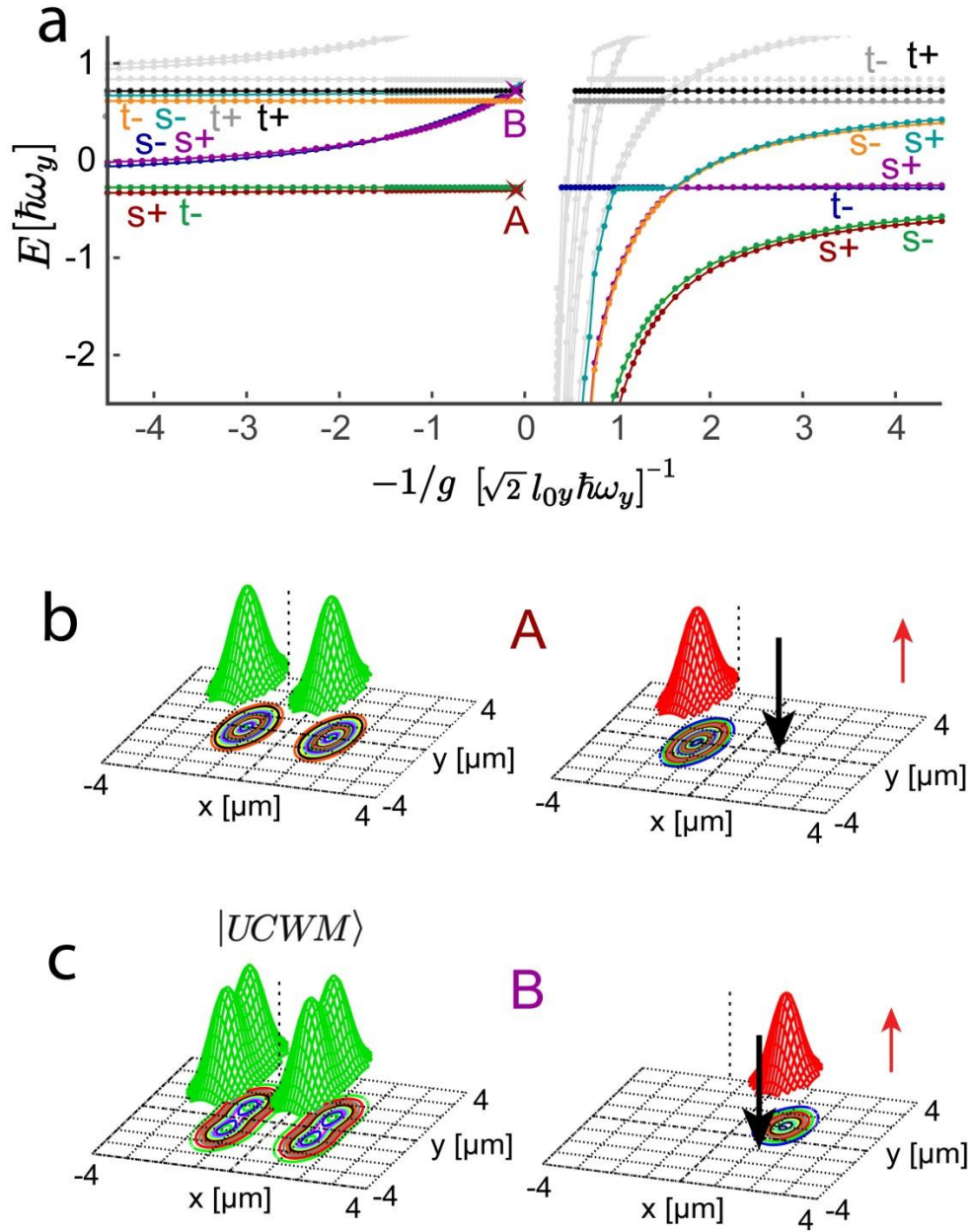

Figure S1

Figure S1. The energy spectrum (a) of two fermions and wavefunction anatomy (b,c) calculated in a double well for the PA configuration. The results are plotted for a configuration with a well separation along the x-axis $d = 2.5\ \mu m = 3.51\ l_0$ ($l_0 \equiv l_{0x} = \sqrt{\hbar/(M\omega_x)} = 713$ nm) and no tilt ($\Delta = 0$), plotted as a function of the inverse interaction strength $-1/g$; $g$ is expressed in units of $\sqrt{2}l_{0y}\hbar\omega_y$, where $l_{0y} = \sqrt{\hbar/(M\omega_y)} = 1297$ nm. The confining frequencies in the $x$ and $y$ directions are $\omega_x = 2\pi \times 6.6$ kHz and $\omega_y = 2\pi \times 1$ kHz. The barrier heights $V_b$ (produced by the smooth neck) are given by $V_b/h = 0.49\ V_0/h = 10kHz$, where $V_0$ is the bare barrier of the TCO double well; $V_0/h = 20.4$ kHz. The mass corresponds to ultracold $^6$Li atoms, $M = 9.99$ $10^{-27}$ kg. The DW parameters are within the range of those used in the experiments[2].

**(a)** Both repulsive $(-1/g < 0)$ and attractive $(-1/g > 0)$ interparticle interactions are considered. The horizontal curves correspond to Heitler –London (HL)-type states (one fermion in each well) that relate to the maximally spin-entangled two-qubit Bell states. Due to parity conservation, the $g$-dependent, doubly-degenerate first-excited (dark blue and violet) energy curves in the repulsive regime correspond to highly space-entangled NOON-type states of the form $(|2,0> \pm |0,2>)/\sqrt{2}$. **(b** and **c)** The many-body wave-function anatomy (single-particle densities, SPD green surfaces, and spin-resolved conditional probability distributions, CPDs, red surfaces) is illustrated for two instances, marked by letters A (shown in (b)) and B (shown in (c)) on the energy curves (in a). The abscissa value associated with these letters is $-1/g = -0.1/(\sqrt{2}l_{0y}\hbar\omega_y)$. The far-left part of the $-1/g$ axis represents the non-interacting limit. Point A (on the s+, positive parity singlet, brown line) is a representative of the above-mentioned HL-type state, and point B (on the s+ , positive parity singlet, purple line) is a representative of a NOON state. In the spin-resolved CPDs (red surfaces) the black down arrow represents the location of the spin-down fermion (taken at one of the humps in the single-particle density plots (green surfaces)), and the red arrow signifies that the red surface corresponds to the up-spin probability distribution. In (b) placing the down-spin fermion at the position of the right well (black down arrow in the position of the right density hump) shows that the distribution of the up-spin fermion (red surface) is found to be located in the other (left) well. The SPD and CPD depicted in (c) are of particular interest, representing a NOON state formed by the superposition of two-fermion ultracold Wigner molecules (UCWMs) located in either the right or left wells. The double-humped SPDs indicate that the two fermions (due to the large repulsion) localize and avoid each other, forming an UCWM. The displayed CPD in (c) confirms formation of a UCWM – indeed, placing the down-spin fermion (black down arrow) at the position of the forward density hump in the right well, the distribution of the up-spin fermion (red surface) is found to be located away from the black arrow with its maximum coinciding with the backward hump in the SPD in the right well. If the fixed (observation) point is chosen to be in the left well, the resulting CPD will be a mirror image of the one shown above, namely it will depict a red surface in the left well. Note that the formation of NOON states is due to the conservation of parity when the detuning tilt ($\Delta$) between the wells vanishes (as is the case here).

**4. Comparison with experiment.** To compare with the experimental results[2] regarding single and double occupancy as a function of the interaction strength, $g$, we first extract from our calculations the relevant Hubbard-model parameters. For the purpose of this comparison we use our calculations for the DW systems in the linear arrangement, LA, and parallel arrangement, PA. For the LA case we use the following parameters (see caption to Fig. 1c in the main text): The confining frequencies in the $x$ and $y$ directions are $\omega_x = 2\pi \times 1$ kHz and $\omega_y = 2\pi \times 100$ kHz, leading to an effective 1D confinement along the $x$ direction. The barrier height $V_b$ (produced by the smooth neck) is $V_b/h = 18.18\ V_0/h = 5.407$ kHz, where $V_0$ is the bare barrier of the TCO double well , $V_0/h = 0.297$ kHz, where h is the Planck constant.. This factor leads to strong anharmonicities in the confining double-trap potential. The interwell separation is $d = 2\ \mu m = 1.543 l_0$, representing two rather well-separated wells, with $l_0 \equiv l_{0x} = \sqrt{\hbar/(M\omega_x)}$ = 1.297 μm being the (left or right) harmonic-oscillator length. The mass corresponds to ultracold $^6$Li atoms, $M = 9.99\ 10^{-27}$ kg. For the PA case we use the parameters given in the caption to Fig. S1. These parameters correspond to those used in the experiment[2], selected there in order to assure applicability of the Hubbard model employed in reference 2, due to the small tunneling (hopping parameter J) between the two wells. .

The Hubbard-model hopping parameter is obtained from the energy spectrum of the non-interacting case for the symmetric double well (with $\Delta = 0$), i.e., the energy difference, 2J, between the singlet ground state and the first-excited triplet state . In this way we extracted for the LA configuration a value of  J/h = 48.73 Hz,  and for the PA configuration J/h = 55.53 Hz, which are sufficiently small compared to the axial trap frequency 1 kHz for the LA  and PA configurations,  corresponding to the weak tunneling regime as in the experiments.

The Hubbard parameter U (the onsite interparticle interaction strength) as a function of $g$ (where $g$ is the contact interaction strength in the microscopic hamiltonian given in Eq. 1 of the main text) is the energy difference, $E(-1/g) - E(-\infty)$, for the singlet ground state in a single well; $E(-\infty)$, is the energy of two non-interacting particle in a single well. If the calculations of U were to be done for the symmetric ($\Delta = 0$) case, the results would contain contributions from interwell tunneling. To minimize interwell tunneling we performed (for both the LA and PA configurations) the above evaluation for U in a strongly tilted DW configuration, so that the low energy spectrum is determined solely by the lower lying well. This also incorporates the effect of anharmonicity which is inherent to the DW confinement; this effect is particularly important in the LA configuration. In these calculations we use a tilt of $\Delta/h = 4.84$ kHz for the LA configuration and $\Delta/h = 9.67$ kHz for the PA configuration, while keeping the other trap parameters unchanged.

Having established the $U(g)$ dependence, we carry out a series of CPD calculations for the *symmetric* double well (with the same $d$ and $V_b$) where the fixed point is placed in the left well ($x < 0$). The portion of the CPD for $x < 0$ yields the probability of double occupancy. On the other hand, the portion of the CPD for $x > 0$ yields the single-occupation probability.

Our calculations compared to the experimental measurements are displayed in Fig. S2.

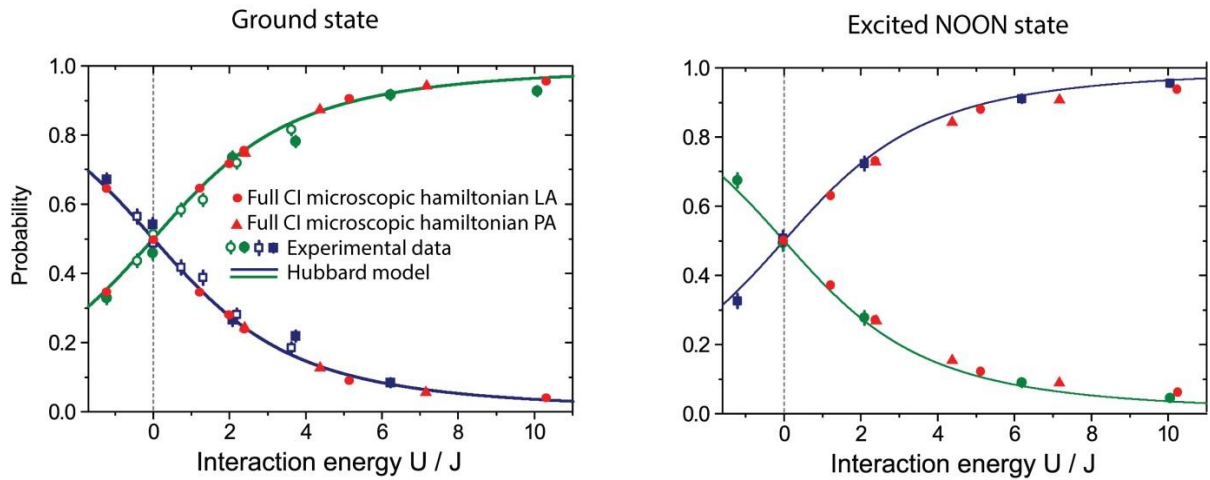

Figure S2

**Figure S2**. **Probability of double (blue curve) and single (green curve) occupations of the left and right wells for the LA and PA double well configurations.** For the ground state

probabilities (left frame) of both configurations we use the s+ singlet ground state (brown curve in Fig. 1c in the main text and in Fig. S1, for the LA and PA, respectively). For the excited NOON state (right frame) we use in the LA configuration the orange s+ singlet of Fig 1c (main text), and in the PA configuration we use the purple s+ singlet curve of Fig. S1. For both DW arrangements we carried out calculations for two repelling $^6$Li atoms in a symmetric ($\Delta = 0$) double well, with the parameters of the calculations described for the LA configuration in the main text (see captions to Fig. 1) and the start of this subsection, and for the PA configuration the parameters are given in the caption of Fig. S1. Blue squares and green circles represent experimental data from Ref. 2. Red circular dots represent our CI simulation results for the LA configuration, and red triangles correspond to our calculated results for the PA configuration. Note the interchange between the blue and green probability curves (compared to the left panel), which is found both in theory and the experiment. Note that the calculated results for both the LA and PA configurations of the double well system agree well with the experimentally measured data. The limit of the Hubbard model cannot distinguish between the two microscopic trap arrangements.